# COMBINATORIAL ON/OFF MODEL FOR OLFACTORY CODING

A.Koulakov[1], A.Gelperin[2], and D.Rinberg[2]

1. Cold Spring Harbor Laboratory, Cold Spring Harbor, NY, 11724 USA

2. Monell Chemical Senses Ctr., Philadelphia, PA, 19104 USA

**We present a model for olfactory coding based on spatial representation of glomerular responses. In this model distinct odorants activate specific subsets of glomeruli, dependent upon the odorant's chemical identity and concentration. The glomerular response specificities are understood statistically, based on experimentally measured distributions of detection thresholds. A simple version of the model, in which glomerular responses are binary (the on/off model), allows us to account quantitatively for the following results of human/rodent olfactory psychophysics: 1) just noticeable differences in the perceived concentration of a single odor (Weber ratios) are $dC/C \sim 0.04$; 2) the number of simultaneously perceived odors can be as high as 12; 3) extensive lesions of the olfactory bulb do not lead to significant changes in detection or discrimination thresholds. We conclude that a combinatorial code based on a binary glomerular response is sufficient to account for the discrimination capacity of the mammalian olfactory system.**

Although the ability of the olfactory system to represent both the quality and intensity of odors has been extensively studied, the olfactory code remains unbroken. One of the main questions in studies of the sense of smell is how the odors are represented in the activity of central olfactory neurons. Experiments on anatomical connectivity[1-4] and imaging of neuronal activity[5-12] provide evidence for the spatial representation of olfactory information in the olfactory bulb (OB)[13-15]. Whether the information about odorants is relayed to higher brain areas in its spatial form or is translated by OB neurons into a more efficient representation, perhaps involving temporal correlations, is unclear. A definitive resolution of this question will allow investigators to directly probe neural representations of olfactory stimuli and ultimately build a testable model for the olfactory code.

Information about odorants in mammals is initially represented in the activity of olfactory sensory neurons (OSNs). Each OSN expresses one and only one type of olfactory receptor (OR) protein[16,17] whose binding specificity makes possible the recognition of odorant molecules. OSNs can therefore be divided into classes expressing distinct OR types. The number of such types ranges from hundreds in humans to thousands in other vertebrates[18,19]. Projections from multiple OSNs expressing the same type of olfactory receptor protein converge to a localized distinct area in the OB, called a glomerulus. Glomeruli relay information about odorants to the mitral/tufted cells, the primary output neurons of OB. Several dozen mitral/tufted cells receive their inputs from one glomerulus and send their outputs to the olfactory cortex and other brain areas.

According to the spatial theory of odor coding olfactory information is represented in the spatial response patterns of glomeruli and their associated mitral/tufted neurons. Just as images on the retina evoke specific patterns of activation of retinal ganglion cells, different odorants are encoded in the spatial pattern of activity of mitral/tufted cells. Although the spatial code is simple it represents a powerful coding scheme, capable of encoding both intensity and quality of different odors, especially if different glomeruli are activated simultaneously in various combinations, leading to the notion of the combinatorial code[20].

As an alternative to the purely spatial coding hypothesis, a temporal theory proposed that olfactory information is represented in the temporal pattern of neuronal spiking and/or in spiking synchronized to collective neuronal oscillations. Some implementations involve information transfer by synchronization of spike timing with respect to the phase of oscillations[21,22]. Although evidence exists for the presence of correlations between mitral/tufted cell activity and local field potential oscillations, it is not clear if this is how olfactory information is primarily transferred. The main argument in favor of the temporal coding theory is that coding in the temporal domain strongly increases the information capacity of the code. However, the information capacity of various types of olfactory codes has not been thoroughly investigated.

In this study we investigate the discrimination capacity of the combinatorial spatial code. The main assumption made is that the elementary unit of the olfactory code is the glomerulus. The mitral cells receiving excitatory inputs within the same glomerulus are assumed to send information into the olfactory cortex about how strongly a given glomerulus is activated. Alternatively, it could be possible that each mitral cell acts as an independent coder, making discrimination of odorant identity possible even within a single glomerulus. This hypothesis will not be pursued here. We will also assume initially that the glomerulus has a binary response and as such can be either active (ON) or inactive (OFF). This assumption is made to simplify the presentation of our results. Later we relax this assumption by allowing graded activation of a single glomerulus.

These assumptions put strong restrictions on the olfactory spatial code and perhaps the only remaining feature of the code is its combinatorial complexity. Having made these assumptions we address results of human and rodent psychophysics. We ask if the observed odorant discrimination capacity observed in these experiments can be explained *only* based on the combinatorial olfactory code. A positive answer to this question will make a strong case for the parsimony of a spatial code. We will address experiments on the human Weber ratio[23], the robustness of olfactory discrimination to lesions observed in rodents[24, 25], and the number of pure odors which can be detected simultaneously[26].

For the purposes of this study we define the glomerulus as a locus in the olfactory bulb which receives inputs from receptor cells expressing the same genetically distinct type of olfactory receptor. Thus two glomeruli receiving inputs from the same receptor neurons in two olfactory bulbs, left and right, will be included in the same independent glomerulus in our model. The number of independent glomeruli defined in this way is equal to the number of genetically distinct types of olfactory receptor proteins expressed by the receptor cells. For humans this number is about $N = 350$ [18, 19]. For mice and rats the number of independent glomeruli is about $N = 1000$ [18].

The glomerular responses in our approach are defined by the set of binary numbers $r_n$, where the index $n$ runs from 1 to $N$. The numbers indicate if a given glomerulus is activated ($r_n = 1$) or not ($r_n = 0$) by the odor. Note that we assume that a glomerulus carries one bit of information, but do not specify how information is transferred to the other part of the brain. Information transfer may be realized in mitral/tufted cells excitatory or inhibitory responses, or just as a deviation from the baseline.

Different odors activate different subsets of glomeruli (Figure 1A) thus resulting in combinatorial encoding of stimulus quality. For increasing concentration of the same odorant, we assume that glomeruli are sequentially recruited, i.e. the number of glomeruli that are active increases with increasing intensity of the stimulus. We therefore assume that no glomerulus can be deactivated by an increased concentration of the same odor (Figure 1B).

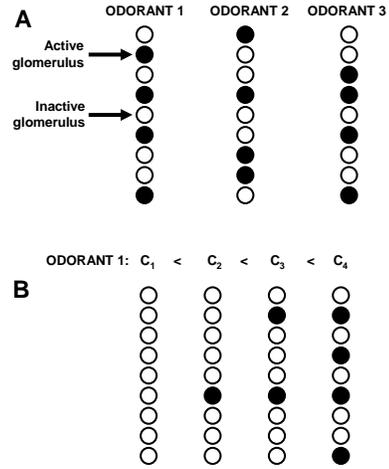

**Figure 1. Responses of independent glomeruli. (A) Different odors activate different subsets of glomeruli. (B) Increase in concentration of the same odorant leads to recruitment of additional glomeruli.**

Our model can be restated in terms of the activation thresholds for the glomeruli. Thresholds $\theta_n(O)$ are defined as the concentrations of the odorant $O$ at which glomerulus number $n$ is activated. For each simple odorant there is a set of $N$ such thresholds, which completely define the response of our model olfactory system to that particular odorant. Indeed, for each value of concentration $C$ the glomeruli satisfying $\theta_n(O) \leq C$ are ON, and vise versa. A possible set of odorant thresholds for the set of glomeruli able to be activated by that odorant is shown in Figure 2. Following Hopfield[27] we will assume that the thresholds are distributed uniformly on the logarithmic scale of concentration. The distribution of thresholds covers about six orders of magnitude in concentration[28]. This value corresponds to $A = 6\ln 10 \approx 14$ in natural logarithm units, which will be used throughout this paper.

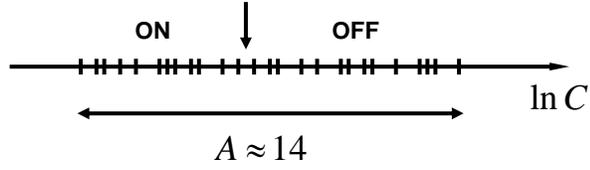

**Figure 2. The thresholds for activation of the set of glomeruli by a simple odorant (vertical marks). For a given odorant concentration (arrow) the glomeruli with smaller thresholds are active (ON). The width of the threshold distribution in the logarithmic scale is $A \approx 14$.**

Our model has therefore two essential numerical parameters: the number of independent glomeruli $N$, which is variable from species to species, and the dynamic range of the olfactory system $A \approx 14$. Using these two parameters we address a variety of psychophysical experiments.

We will first deduce the Weber ratio (the just noticeable relative change in concentration). For a given concentration $C$ the number of recruited glomeruli (in the ON state) is

$$n = N \frac{\ln C}{A}, \quad (1)$$

as evident from Figure 2. The change in concentration is detected if it leads to the recruitment of at least one new glomerulus. Condition $\Delta n = 1$ leads in combination with Eq. (1) to the following expression for the Weber ratio

$$\frac{\Delta C}{C} \approx \frac{A}{N} \quad (2)$$

In other words the Weber ratio is equal to the average distance between neighboring thresholds on the logarithmic scale of concentrations. The estimate for the value of the Weber ratio for humans can be obtained by taking $N = 350$ and $A \approx 14$, which results in $\Delta C / C \approx 4\%$. This result compares favorably with experimental findings of Cain[23], who measured $\Delta C / C \approx 4 \div 16\%$. In particular, when an odorant was delivered using an air-dilution olfactometer, the Weber ratio could reach 4.2% (see Figure 3 in Ref. [[23]]). Values of the Weber ratio larger than 4% are possible in our model if, for example, noise is present in the glomerular response.

We will next examine the effects of lesions of the olfactory bulb on the detection of odors. Experimental studies suggest that extensive bulbar lesions lead to no significant effect in detection and discrimination of odorants[24, 25]. Here we suggest that this conclusion follows naturally from the ON/OFF model. The model therefore reproduces the observed robustness of olfactory discrimination after lesions of the olfactory bulb.

Consider the detection task first. The presence of the odorant is detected in our model if at least one glomerulus is activated. The minimal perceived concentration is therefore determined by the minimum value in the set of $N$ thresholds $\theta_n(O)$ for different glomeruli (Figure 3) responsive to a particular odorant. If the lesion removes a fraction $f$ of all glomeruli, chosen randomly, the minimum threshold shifts to the next lowest available threshold, which is spared by the lesion. The average shift in the detection threshold is given by

$$\frac{\Delta C_{Min}}{C_{Min}} = \frac{A}{N} \frac{f}{1-f} \quad (3)$$

For a 50% bulbar lesion (50% of glomeruli are removed, $f = 0.5$) it coincides with the Weber ratio (2), which for rats is equal to $\Delta C / C = 0.014$ in this model ($N = 1000$). The shift in the detection threshold given by equation (3) is indeed insignificant, which renders the olfactory system robust to lesions. The robustness stems from the broad tuning of different glomeruli: If some of them are removed, others can still detect an odorant.

The ability to detect a given odorant also implies that two different odorants can be discriminated in this model. Indeed, if each odorant is presented at the perceptual threshold, it will activate a single glomerulus. By detecting which glomerulus is active one could infer what odorant is present. If more than one glomerulus is activated by each odorant, discrimination becomes more reliable. Thus, both detection and discrimination are robust to lesions in this model.

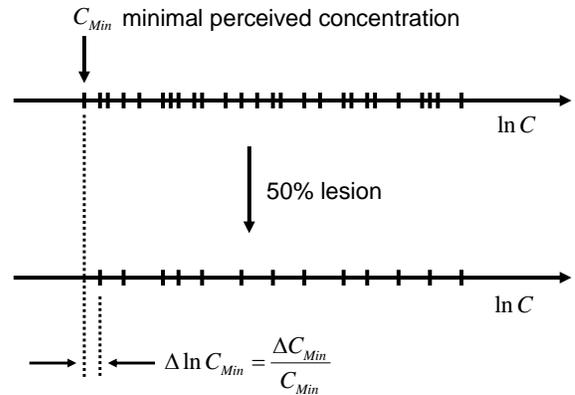

**Figure 3. The robustness of a combinatorial code to lesions. The detection threshold for a given odorant is determined by the minimum threshold**

**out of the set of $N$ different glomeruli potentially activated by that odorant. A lesion of the olfactory bulb removes a subset of the glomeruli. The detection threshold in the lesioned bulb is determined by the next lowest available threshold, which survives the lesion. The shift in the detection threshold due to a lesion is small and is of the order of the Weber ratio.**

We will now address the number of pure odorants that can be identified simultaneously in our model. The basic problem with identifying pure odorants in the odor mixture is illustrated in Figure 4. Assume that the subject is presented with a mixture of two odorants: O1 and O2. This mixture is identified as M1 in Figure 4. A possible response of the glomeruli to the mixture includes a union of the glomerular activation evoked by O1 and O2 when presented separately. The rules for addition of responses to two or more components in this model correspond to a logical OR function between these components, reproducing the behavior of inputs from hypoadditive receptor neurons [see e.g. Duchamp-Viret et al, 2003 [29]]. When the subject is presented with another mixture M2, in which O2 is replaced by O3, the pattern of activation may be exactly the same as in the response to M1. In this event O2 cannot be distinguished reliably from O3 in the mixture. The quantitative question that arises in our model is at what number of pure odorants in the mixture such ambiguity may arise.

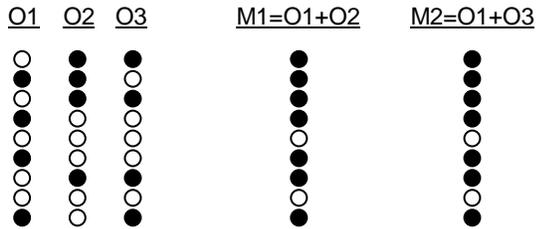

**Figure 4. Responses to mixtures in the ON/OFF model. Three pure odorants mixed in different combinations (M1 and M2) evoke the same response in the glomerular array, leading to the failure to identify the presence of the individual components of the mixture.**

To identify the number of pure odorants at which the ON/OFF model fails to differentiate the presence of one of them in a mixture we notice the following condition in Figure 4. If one adds O2 to mixture M2, the glomerular response is unchanged. This implies that *no* glomeruli are recruited by adding an extra odor to the mixture. The failure of the organism to identify an extra odor occurs when $\Delta g < 1$, where $\Delta g$ is the average number of glomeruli recruited by the new odor.

Consider a mixture of $S$ odorants. Our goal is to determine the number $g(S)$ of glomeruli activated by this mixture on average. To accomplish this goal we will relate $g(S)$ to the number of glomeruli $g(S+1)$ active when another component, called O, is added to the mixture. Assume that the component O, when presented alone, activates $n$ glomeruli. In the presence of other odorants the number of newly recruited by component O glomeruli is expected to be smaller than $n$. Indeed, additional components of the mixture activate a fraction $x = g(S)/N < 1$ of all glomeruli, and $xn$ of the glomeruli recruited by the additional component O. These glomeruli are already activated and cannot be recruited by O. Therefore one expects that the number of glomeruli newly recruited by O is $n - xn$ rather than $n$. The equation that determines the rate of recruitment when new components are sequentially added to the mixture is

$$g(S+1) = g(S) + n - xn. \qquad (4)$$

Using $x = g(S)/N$ this equation can be solved to result in the desired number of glomeruli recruited when $S$ component mixture is presented

$$g(S) = N\left[1 - \left(1 - \frac{n}{N}\right)^S\right]. \qquad (5)$$

Here $n$ is the average number of glomeruli activated by a single pure odorant, given by equation (1). This number is assumed to be similar for different components. A more general equation, which includes different numbers of glomeruli activated by each component $n_s$, is

$$g(S) = N\left[1 - \prod_{s=1}^{S}\left(1 - \frac{n_s}{N}\right)\right]. \qquad (6)$$

Here the product is assumed over all components present in the mixture.

As we mentioned, the olfactory system described here fails to detect a substitution of one of the odors if $\Delta g \equiv g(S+1) - g(S) < 1$. Using (5) we obtain from this condition the maximum number of odorants in the mixture which can be detected

$$S^* = \frac{\ln n}{\ln\left[N/(N-n)\right]} \qquad (7)$$

Equation (7) is illustrated in Figure 5A (solid line).

We will now examine the maximum number of detectable components in the intermediate range of concentrations. In this case, as follows from equation (1), single components activate a small fraction of the bulb, i.e. $n \ll N$. In this regime the maximum number of components (7) becomes

$$S^* = \frac{N}{n}\ln n = A\frac{\ln\left(A^{-1}N\ln C\right)}{\ln C}. \quad (8)$$

The magnitude of the maximum number of odorants is therefore determined by the dynamic range of the natural logarithm of distinguishable concentrations $A \approx 14$, which corresponds to about six orders of magnitude of concentration.

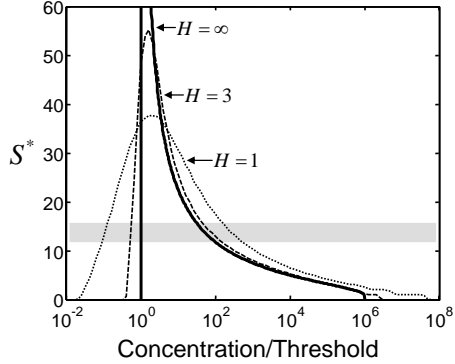

**Figure 5. The average number of components in a mixture that permits detection of a single component. Solid line shows the results for the ON/OFF model [equation (7)]. The results for the models involving graded glomerular activation are shown by the dashed (Hill coefficient $H = 3$) and dotted $H = 1$ lines for comparison. The experimentally observed range of values is shown by the gray region.**

Experimentally the number of pure odorants in mixture at which the presence or absence of a single component can be detected is between 12 and 15 [26]. This number is consistent with the concentration of individual components to be about $10^2$ of threshold for detection (Figure 5). Our model predicts that with increasing concentration of individual odorants, the number of detectable components should decrease. The overall order of magnitude of the number of odorants detectable in a mixture is determined by the natural logarithm of the range of concentrations distinguished by the olfactory system, which is given by the parameter $A \approx 14$ [see equation (8)].

We now will address some of the assumptions made in this work. In particular, we would like to find the conditions under which glomerular responses can be considered binary. To account for the graded nature of the response of a glomerulus number $i$ as a function of concentration $r_i(C)$ we assume that it is described by the Hill equation

$$r_i(C) = \frac{C^H}{C^H + K_i^H} = \frac{1}{1+e^{H(\ln K_i - \ln C)}} \quad (9)$$

Here $H$ and $K_i$ are the Hill exponent and the saturation concentration, respectively, for this glomerulus. We normalized the response by the maximum value to make a comparison with the ON/OFF case. The second equation in (9) emphasizes that as a function of the logarithm of concentration the Hill equation actually takes the form of a logistic function. The steepness of the logistic function is determined by the Hill exponent: The response increases from 0 to 1 within the range of the logarithm of concentration proportional to $1/H$. As the Hill exponent increases the glomerular response becomes sharper, until, in the limit $H \to \infty$, it becomes infinitely sharp, as in the ON/OFF model considered above. At what value of the Hill exponent can one consider the ON/OFF model to be valid?

To make a connection to the ON/OFF model we construct the integrated population response, which represents the number of active glomeruli in the case of graded responses[11, 12, 30]

$$n = \sum_{i=1}^{N} r_i(C) \quad (10)$$

The glomeruli whose saturation concentration $K$ is far below $C$ contribute unity to the sum, thus playing the role of ON units. Glomeruli, for which $C$ is much smaller than the threshold, contribute little to the population activity playing the role of OFF units (Figure 6). The graded contribution between 0 and 1 is expected for glomeruli in the range of $\ln K$ of width $1/H$ around the current value of concentration (Figure 6). The saturation concentration $K$ thus plays the role of threshold concentration for glomerular activation. When sum (10) is evaluated for the pure ON/OFF model ($H \to \infty$), it renders the number of active glomeruli (1). If one considers the case of $H \sim 1$ one notices that the integral population activity (10) does not differ much from the ON/OFF case ($H \to \infty$). This is because the contribution from superthreshold glomeruli is lowered (Figure 6), while the glomeruli with the thresholds lower than the odorant concentration contribute more, leading to compensation and no substantial change in the integral population activity due to the finite Hill coefficient. This cancellation is possible if the range of thresholds for glomerular activation $A$ is substantially larger than the range of graded response for a single glomerulus $1/H$:

$$A \gg 1/H \quad (11)$$

In other words, the dynamic range for the entire population should exceed substantially the dynamic range for a single detector. In this case the graded response of a single detector (glomerulus) becomes irrelevant and one should rely on the population activity to use the entire dynamic range. If this condition is satisfied the ON/OFF model captures the essential behavior of the population activity. For the activation thresholds distributed within six orders of magnitude in concentration ($A \approx 14$) and the Hill coefficients ranging between about 0.5 and 4.4 [12, 30] one expects condition (11) to be met and the validity of the ON/OFF model to be satisfied.

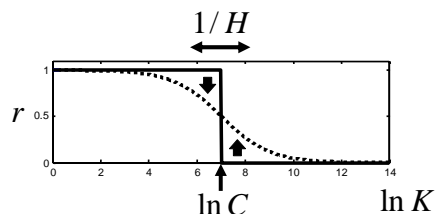

**Figure 6. The response of a population of glomeruli to an odorant at concentration $C$ as a function of the saturation concentration $K$. Solid line: responses in the ON/OFF model (Hill coefficient $H \to \infty$). Dotted line: the responses for $H = 1$. The total integrated response of the population is the same for dotted and solid lines. This is because changes in response are of opposite sign for glomeruli whose saturation concentrations are above and below the odorant concentration (up and down arrows). For this reason the ON/OFF model is valid to describe population response when condition (11) is satisfied.**

We also examined the effect of graded glomerular responses on the number of components in a mixture that can be detected. To this end we evaluated the change in the response of the glomerular array caused by the replacement of a single component. The response to individual components is described by Eqs. (9) and (10). As in the ON/OFF case we assumed that the activation pattern for mixture is given by the maximum over the responses obtained for individual components, that corresponds to hypoadditivity[29]. We also assumed that substitution of a component can be detected, if the total variation of the glomerular response vector is equal to one, similar to the pure ON/OFF case. The critical number of components discriminated in a mixture obtained under this condition is shown in Figure 5 by the dashed ($H = 3$) and dotted ($H = 1$) lines. The main effect of graded glomerular responses is in smearing the ON/OFF dependence by the amount $\sim 1/H$. For moderate concentrations of components the analytical result (7) gives an adequate approximation to the results obtained in the case of graded glomerular responses. We conclude that the ON/OFF model may give a numerically exact representation of the number of components in the mixture that can be detected even if the actual responses of glomeruli are graded. Also, except for the small range of the values of components' concentration around threshold for detection, the ON/OFF model underestimated the number of components that can be detected (Figure 5). Therefore additional information about the mixture composition was provided by the graded responses of glomeruli.

In this work we examine a simplified model for olfactory coding involving ON/OFF glomeruli. The discrimination between odor concentration and composition in this model is possible by examining binary strings, representing glomerular response patterns. Our study therefore examines the purely combinatorial component of the olfactory code. We conclude that the results of psychophysical studies probing the discrimination capacity of human and rodent olfactory systems can be understood on the basis of this simplified model. In some cases, such as human Weber ratios, the simplified model predicts a better performance than displayed by humans. Our argument however is not that human/rodent olfactory systems employ a binary code. Instead we suggest that a spatial combinatorial code is sufficient to explain the discrimination capacity of human/rodent olfaction.

Additional features can be added to our model for a spatial code, such as graded glomerular responses or non-linear interactions between mixture components. In this case the olfactory code is expected to become more powerful and the discrimination capacity should increase in the broad range of parameters. Our estimates therefore provide lower bounds for the discrimination capacity of the spatial code. Since these lower bounds appear to be in good agreement with experiments, the spatial code provides both a simple and powerful scheme for representing olfactory information.